\documentclass[11pt]{article}
\usepackage{graphicx}

\newcommand{\BABARPubYear}    {00}
\newcommand{\BABARPubNumber}  {12}
\newcommand{\SLACPubNumber} {8534}

\usepackage{relsize}
\def\babar{\mbox{\slshape B\kern-0.1em{\smaller A}\kern-0.1em
    B\kern-0.1em{\smaller A\kern-0.2em R}}}

\def\en         {\ensuremath{e^-}}      
\def\ep         {\ensuremath{e^+}}
  
\def\epem       {\ensuremath{e^+e^-}}

\def\qqbar {\ensuremath{q\overline q}}

\def\piz   {\ensuremath{\pi^0}}

\def\pip   {\ensuremath{\pi^+}}
\def\pim   {\ensuremath{\pi^-}}

\def\Kbar  {\ensuremath{\kern 0.2em\overline{\kern -0.2em K}}}

\def\Kp    {\ensuremath{K^+}}
\def\Km    {\ensuremath{K^-}}

\def\Kstarz  {\ensuremath{K^{*0}}}

\def\Kstarb   {\ensuremath{{\Kbar}\kern0.03em {\raise.18ex\hbox{$^*$}}}}

\def\Kzb   {\ensuremath{{\Kbar}\kern0.03em {\raise.1ex\hbox{$^0$}}}}
\def\KzKzb {\ensuremath{K^0 {\kern -0.16em \Kzb}}}
\def\Dz    {\ensuremath{D^0}}
\def\Dbar  {\ensuremath{\kern 0.2em\overline{\kern -0.2em D}}}
\def\Dzb   {\ensuremath{{\Dbar}\kern0.03em {\raise.3ex\hbox{$^0$}}}}
\def\DzDzb {\ensuremath{D^0 {\kern -0.16em \Dzb}}}

\def\Dstarb   {\ensuremath{{\Dbar}\kern0.03em {\raise.18ex\hbox{$^*$}}}}

\def\Bz    {\ensuremath{B^0}}

\def\Bbar  {\ensuremath{\kern 0.18em\overline{\kern -0.18em B}}}
\def\Bzb   {\ensuremath{{\Bbar}\kern0.03em {\raise.3ex\hbox{$^0$}}}}

\def\BB    {\ensuremath{B\Bbar}} 
\def\BzBzb {\ensuremath{B^0 {\kern -0.16em \Bzb}}}

\mathchardef\Upsilon="7107
\def\Y#1S{\ensuremath{\Upsilon{(#1S)}}}

\def\FourS {\Y4S}

\mathchardef\Delta="7101
\mathchardef\Xi="7104
\mathchardef\Lambda="7103
\mathchardef\Sigma="7106
\mathchardef\Omega="710A
\def\Deltabar   {\ensuremath{\kern 0.25em\overline{\kern -0.25em \Delta}}{}}
\def\Lbar {\ensuremath{\kern 0.2em\overline{\kern -0.2em\Lambda\kern 0.05em}\kern-0.05em}{}}
\def\Sigbar{\ensuremath{\kern 0.2em\overline{\kern -0.2em \Sigma}}{}}
\def\Xibar{\ensuremath{\kern 0.2em\overline{\kern -0.2em \Xi}}{}}
\def\Obar{\ensuremath{\kern 0.2em\overline{\kern -0.2em \Omega}}{}}
\def\Nbar{\ensuremath{\kern 0.2em\overline{\kern -0.2em N}}{}}

\def\BR{{\ensuremath{\cal B}}}

\def\btosgam   {\ensuremath{b \to s \gamma}}
\def\btodgam   {\ensuremath{b \to d \gamma}}

\def\pt         {\mbox{$p_T$}}
\def\mes        {\mbox{$m_{\rm ES}$}}

\def\ev   {\ensuremath{\rm \,e\kern -0.08em V}}
\def\kev  {\ensuremath{\rm \,ke\kern -0.08em V}} 
\def\mev  {\ensuremath{\rm \,Me\kern -0.08em V}} 
\def\gev  {\ensuremath{\rm \,Ge\kern -0.08em V}} 
\def\gevc {\ensuremath{{\rm \,Ge\kern -0.08em V\!/}c}} 
\def\tev  {\ensuremath{\rm \,Te\kern -0.08em V}}
\def\mevc {\ensuremath{{\rm \,Me\kern -0.08em V\!/}c}} 
\def\gevcc{\ensuremath{{\rm \,Ge\kern -0.08em V\!/}c^2}} 
\def\mevcc{\ensuremath{{\rm \,Me\kern -0.08em V\!/}c^2}}

\def\invfb   {\ensuremath{\mbox{\,fb}^{-1}}}
\def\mus  {\ensuremath{\rm \,\mus}}

\def\mus        {\ensuremath{\,\mu{\rm s}}}    
\def\gsim{{~\raise.15em\hbox{$>$}\kern-.85em
          \lower.35em\hbox{$\sim$}~}}
\def\lsim{{~\raise.15em\hbox{$<$}\kern-.85em
          \lower.35em\hbox{$\sim$}~}}

\def\to                 {\ensuremath{\rightarrow}}

\def\pep2{PEP-II}

\def\chic1{\ensuremath{\chi_{c1}}}
\def\chic2{\ensuremath{\chi_{c2}}}
\def\chic3{\ensuremath{\chi_{c3}}}

\newcommand{\eqref}[1]{Eq.~(\ref{eq:#1})}

\newcommand{\epjc}      [1]  {{Eur.\ Phys.\ Jour.\ C~{\bf #1}}}

\newcommand{\nim}       [1]  {{Nucl.\ Instr.\ and Methods~{\bf #1}}}
\newcommand{\nima}      [1]  {{Nucl.\ Instr.\ Methods {\bf A{\bf #1}}}}

\newcommand{\npb}       [1]  {{Nucl.\ Phys.\ {\bf B{\bf #1}}}}

\newcommand{\npbps}     [1]  {{Nucl.\ Phys.\ B Proc.\ Suppl.\ {\bf #1}}}

\newcommand{\pl}        [1]  {{Phys.\ Lett.\ {\bf #1}}}      

\newcommand{\prl}       [1]  {{Phys.\ Rev.\ Lett.\ {\bf #1}}} 
\newcommand{\pr}        [1]  {{Phys.\ Rev.\ {\bf #1}}}
\newcommand{\progtp}    [1]  {{Prog.\ Th.\ Phys.\ {\bf #1}}}
\newcommand{\rmp}       [1]  {{Rev.\ Mod.\ Phys.\ {\bf #1}}}  

\newcommand{\zpc}       [1]  {{Z.\ Phys.\ C~{\bf #1}}}
\newcommand{\zp}        [1]  {{Z.\ Phys.\ {\bf #1}}}

\newcommand{\mpl}     [1]  {{Mod.\ Phys.\ Lett.\ {\bf #1}}}

\def\jetset74   {\mbox{\tt Jetset \hspace{-0.5em}7.\hspace{-0.2em}4}}

\def\bkog    {\ensuremath {\Bz \to \Kstarz \gamma}}
\def\kpi    {\ensuremath {\Kstarz \to \Kp \pim}}
\def\brog    {\ensuremath {\Bz \to \rho \gamma}}

\def\eerad    {\ensuremath { \epem \to \epem \gamma}}
\def\taurho {\ensuremath { \tau^{-} \to \rho^{-} \nu, \rho^{-} \to \pim \piz}}

\def\tg     {\ensuremath {\theta^{*}_T}}
\def\ctg     {\ensuremath {\cos{\tg}}}
\def\cth     {\ensuremath {\cos{\theta^{*}_{H}}}}
\def\ctb     {\ensuremath {\cos{\theta^{*}_{B}}}}

\def\ebeam     {\ensuremath {E^{*}_{b}}}
\def\egcms     {\ensuremath {E^{*}_{\gamma}}}
\def\de        {\ensuremath {\Delta E^{*}}}
\def\mkpi      {\ensuremath {M_{\Kp \pim}}}
\def\mpl #1 #2 #3 {Mod.~Phys.~Lett.~{\bf#1},\ #2 (#3)}
\def\npb  #1 #2 #3 {Nucl.~Phys.~B~{\bf#1},\ #2 (#3)}
\def\plb  #1 #2 #3 {Phys.~Lett.~B~{\bf#1},\ #2 (#3)}
\def\pr   #1 #2 #3 {Phys.~Rep.~{\bf#1},\ #2 (#3)}
\def\prd  #1 #2 #3 {Phys.~Rev.~D~{\bf#1},\ #2 (#3)}
\def\prl  #1 #2 #3 {Phys.~Rev.~Lett.~{\bf#1},\ #2 (#3)}
\def\RMP  #1 #2 #3 {Rev.~Mod.~Phys.~{\bf#1},\ #2 (#3)}
\def\zpc  #1 #2 #3 {Z.~Phys.~C~{\bf#1},\ #2 (#3)}
\def\nim  #1 #2 #3 {Nucl.~Instrum.~Methods~{\bf#1},\ #2 (#3)}
\def\nima  #1 #2 #3 {Nucl.~Instrum.~Methods~A.{\bf#1},\ #2 (#3)}
\def\epjc #1 #2 #3 {Euro.~Phys.~Jour~{\bf#1},\ #2 (#3)}
\def\rmp #1 #2 #3 {Rev.~Mod.~Phys~{\bf#1},\ #2 (#3)}
\def\npbps #1 #2 #3 {Nucl.~Phys.~B.~proc.~suppl~{\bf#1},\ #2 (#3)}
\def\progtp #1 #2 #3 {Prog.~Theo.~Phys~{\bf#1},\ #2 (#3)}
\def\etal{{\it et al.}}

\def\brresult    {\ensuremath {\BR(\bkog)= (5.4 \pm 0.8 \pm 0.5) \times 10^{-5}}}
\def\lumi    {\ensuremath {(8.6 \pm 0.3) \times 10^6}}

\gdef\Feynmanlength{\setlength{\unitlength}{0.01pt}}  

\global\newcount\LINETYPE                     
\global\newcount\LINEDIRECTION
\global\newcount\LINECONFIGURATION
\newcommand{\LTYPE}{\LINETYPE}
\newcommand{\LDIR}{\LINEDIRECTION}

\global\LINETYPE=1  \global\LINEDIRECTION=0  \global\LINECONFIGURATION=0
\global\newcount\fermion    \fermion=1
\global\newcount\scalar     \scalar=2
\global\newcount\photon     \photon=3
\global\newcount\gluon      \gluon=4
\global\newcount\especial   \especial=5
\gdef\N{0}  \gdef\NE{1}  \gdef\E{2}   \gdef\SE{3}
    \gdef\W{6}   
\global\newcount\REG            \global\REG=0
\global\newcount\FLIPPED        \global\FLIPPED=1
\global\newcount\CURLY          \global\CURLY=2
\global\newcount\FLIPPEDCURLY   \global\FLIPPEDCURLY=3
\global\newcount\FLAT           \global\FLAT=4
\global\newcount\FLIPPEDFLAT    \global\FLIPPEDFLAT=5
\global\newcount\CENTRAL        \global\CENTRAL=6
\global\newcount\FLIPPEDCENTRAL \global\FLIPPEDCENTRAL=7
             
\global\newcount\SQUASHEDGLUON  \global\SQUASHEDGLUON=8

\newcount\adjx \adjx=0
\newcount\adjy \adjy=0
\global\newdimen\BIGPHOTONS     \BIGPHOTONS=0pt  
\global\newdimen\THICKPHOTONS     \THICKPHOTONS=0pt  
\global\newdimen\THICKPHOTONSWITCH    \THICKPHOTONSWITCH=0pt
\gdef\THICKPHOTONTEST{
\THICKPHOTONSWITCH=0pt
\ifdim\THICKPHOTONS=0pt \relax
  \else \ifnum\LTYPE=3
           \ifnum\LDIR=2 \THICKPHOTONSWITCH=1pt \fi 
           \ifnum\LDIR=6 \THICKPHOTONSWITCH=1pt \fi 
        \fi
\fi
}  

\global\newcount\phantomswitch   \global\phantomswitch=0
\global\newcount\stemlength   \global\stemlength=275   
\global\newcount\absstemlength        
\global\newcount\stemlengthx          
\global\newcount\stemlengthy          
\newdimen\FRONTSTEM  \FRONTSTEM=0pt   
\newdimen\BACKSTEM   \BACKSTEM=0pt    
\newdimen\EITHERSTEM \EITHERSTEM=0pt  
\global\newcount\arrowlength                
\global\newdimen\ATTIP   \global\ATTIP=0pt  
\global\newdimen\ATBASE  \global\ATBASE=1pt 
\global\newcount\unitboxnumber  
\global\newcount\unitboxnumberpo  
\global\newcount\particlelengthx  
\gdef\plengthx{\particlelengthx}
\global\newcount\particlelengthy  
\gdef\plengthy{\particlelengthy}  
\global\newcount\boxlengthx  
\global\newcount\boxlengthy  
\global\newcount\particleadjustx  
\global\newcount\particleadjusty  
\global\newcount\particlelength   
\global\newcount\particlefrontx
\gdef\pfrontx{\particlefrontx}
\global\newcount\PFRONTx
\global\newcount\particlefronty
\gdef\pfronty{\particlefronty}
\global\newcount\PFRONTy
\global\newcount\particlebackx
\gdef\pbackx{\particlebackx}
\global\newcount\particlebacky
\gdef\pbacky{\particlebacky}
\global\newcount\particlemidx
\gdef\pmidx{\particlemidx}
\global\newcount\particlemidy
\gdef\pmidy{\particlemidy}
\global\newcount\seglength  \global\newcount\gaplength
\global\gaplength=850  
\global\seglength=1416  
\global\newcount\Xone    \global\newcount\Yone    
\global\newcount\Xtwo    \global\newcount\Ytwo    
\global\newcount\Xthree  \global\newcount\Ythree  
\global\newcount\Xfour   \global\newcount\Yfour   
\global\newcount\Xfive   \global\newcount\Yfive   
\global\newcount\Xsix    \global\newcount\Ysix    
\global\newcount\Xseven  \global\newcount\Yseven  
\global\newcount\Xeight  \global\newcount\Yeight  
\newsavebox{\lastline}  
\global\newcount\numlineparts   
\global\newcount\upperlineadjx  \upperlineadjx=0  
\global\newcount\upperlineadjy  \upperlineadjy=0  
\global\newcount\lowerlineadjx  \lowerlineadjx=0  
\global\newcount\lowerlineadjy  \lowerlineadjy=0  
\global\newcount\thirdlineadjx  \thirdlineadjx=0  
\global\newcount\thirdlineadjy  \thirdlineadjy=0  
\global\newcount\fourthlineadjx \fourthlineadjx=0  
\global\newcount\fourthlineadjy \fourthlineadjy=0  
\global\newcount\unitboxwidth   \unitboxwidth=1000
\global\newcount\unitboxheight  \unitboxheight=0  
\global\newcount\numupperunits  \numupperunits=8  
\global\newcount\numlowerunits  \numlowerunits=8  
\global\newcount\numthirdunits  \numthirdunits=8  
\global\newcount\numfourthunits \numfourthunits=8  
\global\newcount\fermioncount   \global\fermioncount=0    
\global\newcount\scalarcount    \global\scalarcount=0    
\global\newcount\photoncount    \global\photoncount=0    
\global\newcount\gluoncount     \global\gluoncount=0    
\global\newcount\especialcount  \global\especialcount=0    
\global\newcount\vertexcount    \global\vertexcount=-1
\global\newcount\XDIR
\global\newcount\YDIR
\gdef\SETDIR{  
\ifcase\LDIR 
     \global\XDIR=0  \global\YDIR=1   
\or  \global\XDIR=1  \global\YDIR=1   
\or  \global\XDIR=1  \global\YDIR=0   
\or  \global\XDIR=1  \global\YDIR=-1  
\or  \global\XDIR=0  \global\YDIR=-1  
\or  \global\XDIR=-1 \global\YDIR=-1  
\or  \global\XDIR=-1 \global\YDIR=0   
\or  \global\XDIR=-1 \global\YDIR=1   
\else\DIRECTERROR 
\fi}  
\gdef\moduloeight#1{
\ifnum#1>7 \global\advance #1 by -8 
\relax
\moduloeight#1 
\relax
\else \relax  
\fi}
\gdef\multroothalf#1{\global\multiply #1 by 7071 \global\divide #1 by 10000}
\gdef\negate#1{\global\multiply #1 by -1}

\gdef\slanttest(#1,#2){ 
\ifodd\LDIR
\multiply #1 by 7071  \divide #1 by 10000
\multiply #2 by 7071  \divide #2 by 10000
\fi
}
\gdef\gslanttest(#1,#2){
\ifodd\LDIR
\multroothalf#1
\multroothalf#2
\fi
}
\gdef\setplength{ 
\global\particlelengthx=\unitboxwidth
\global\particlelengthy=\unitboxheight
\global\multiply \particlelengthx by \unitboxnumber
\global\multiply \particlelengthy by \unitboxnumber
\global\advance \particlelengthx by \particleadjustx
\global\advance \particlelengthy by \particleadjusty
}
\gdef\boxlengthdefault{  
\global\boxlengthx=\plengthx
\global\boxlengthy=\plengthy
\ifnum\plengthx<0 \global\multiply\boxlengthx by -1 \fi
\ifnum\plengthy<0 \global\multiply\boxlengthy by -1 \fi
}
\gdef\rearcoords{  
\global\particlebacky=\particlefronty 
\global\particlebackx=\particlefrontx 
\global\advance \particlebackx by \particlelengthx
\global\advance \particlebacky by \particlelengthy
}
\gdef\midcoords{  
\global\particlemidy=\particlefronty
\global\particlemidx=\particlefrontx
\global\stemlengthx=\particlelengthx  
\global\stemlengthy=\particlelengthy  
\global\divide\stemlengthx by 2
\global\divide\stemlengthy by 2
\global\advance \particlemidx by \stemlengthx
\global\advance \particlemidy by \stemlengthy
}
\gdef\setparticle{\setplength\rearcoords\midcoords\boxlengthdefault}  
\gdef\setcoords(#1,#2,#3)(#4,#5,#6)[#7,#8]{  
\global\upperlineadjx=#1
\global\lowerlineadjx=#2
\global\thirdlineadjx=#3
\global\upperlineadjy=#4
\global\lowerlineadjy=#5
\global\thirdlineadjy=#6
\global\unitboxwidth=#7
\global\unitboxheight=#8
}
\gdef\drawoldpic#1(#2,#3){  
\global\particlefrontx=#2
\global\particlefronty=#3
\rearcoords  
\midcoords
\put(#2,#3){\usebox{#1}}
}
\gdef\drawsavedline`#1' as #2[#3#4](#5,#6)[#7]{
\global\LINETYPE=#2
\global\LINEDIRECTION=#3
\global\LINECONFIGURATION=#4
\global\particlefrontx=#5
\global\particlefronty=#6
\global\unitboxnumber=#7  
\selectcase
\rearcoords
\midcoords
\ifnum\phantomswitch=0 \drawas{#1}\fi
}

\gdef\drawas#1{
\global\savebox{#1}(\boxlengthx,\boxlengthy){
\setlength{\unitlength}{0.01pt}
\begin{picture}(\boxlengthx,\boxlengthy)
\multiput(\upperlineadjx,\upperlineadjy)(\unitboxwidth,\unitboxheight)
{\numupperunits}{\upperunitbox}
\ifnum\numlineparts > 1  
\multiput(\lowerlineadjx,\lowerlineadjy)(\unitboxwidth,\unitboxheight)
{\numlowerunits}{\lowerunitbox}  
\fi
\ifnum\numlineparts > 2  
\multiput(\thirdlineadjx,\thirdlineadjy)(\unitboxwidth,\unitboxheight)
{\numthirdunits}{\thirdunitbox}  
\fi
\ifnum\numlineparts > 3  
\multiput(\fourthlineadjx,\fourthlineadjy)(\unitboxwidth,\unitboxheight)
{\numfourthunits}{\lowerunitbox}  
\fi
\end{picture} }
\global\PFRONTx=\pfrontx  \global\PFRONTy=\pfronty   
\SETFRONTSTEM
\THICKPHOTONTEST
\ifdim\THICKPHOTONSWITCH=1pt\global\advance\PFRONTy by 20  \fi
\put(\PFRONTx,\PFRONTy) {\usebox{#1}}   
\ifdim\THICKPHOTONSWITCH=1pt
\global\advance\PFRONTy by -40
\put(\PFRONTx,\PFRONTy) {\usebox{#1}}   
\global\advance \PFRONTy by 20  
\fi  
\SETBACKSTEM
\seglength=1416   \gaplength=850   
}
\gdef\drawandsaveline`#1' as #2[#3#4](#5,#6)[#7]{
\global\newsavebox{#1}
\drawsavedline`#1' as #2[#3#4](#5,#6)[#7]
}

\gdef\drawline#1[#2#3](#4,#5)[#6]{   
\drawsavedline`\lastline' as #1[#2#3](#4,#5)[#6]}

\gdef\TYPEERROR{\message{*** ERROR IN PARTICLE TYPE SELECTION ***}
\message{+++ Try with line type \fermion,\scalar,\photon,\gluon 
(see manual) +++}\SETERR}
\gdef\DIRECTERROR{\SETERR\message{*** ERROR IN PARTICLE DIRECTION SELECTION ***}
\message{+++ Try again with direction N, NE, E, SE  etc. or see manual +++}}
\gdef\UNIMPERROR{\message{*** ERROR IN PARTICLE OPTIONS SELECTION ***}
\message{
+++ The requested options combination has not yet been implemented +++}\SETERR}
\gdef\SETERR{\gdef\upperunitbox{{\tiny Error}}  
\gdef\lowerunitbox{\relax}
\gdef\thirdunitbox{\relax}
}
\gdef\neglengthcheck{\ifnum\unitboxnumber < 1 
\message{   *** ERROR:  PARTICLE OF NEGATIVE OR ZERO LENGTH REQUESTED. ***   }
\message{   ***         TAKING ABSOLUTE VALUE. ***   }\negate\unitboxnumber \fi}
\gdef\selectcase{  
\neglengthcheck   
\SETDIR  
\ifcase\LINETYPE
\TYPEERROR  
\or \selectfermion  
\or \selectscalar   
\or \selectphoton   
\or \selectgluon    
\or \selectespecial 
\else \TYPEERROR \fi  }
\gdef\selectfermion{
\ifnum\fermioncount=0 
\global\newcount\fermionlength  
\global\newcount\fermionlengthx
\global\newcount\fermionlengthy
\global\newcount\fermionfrontx  
\global\newcount\fermionfronty  
\global\newcount\fermionbackx
\global\newcount\fermionbacky
\gdef\ALLfermion{  
\global\fermionfrontx=\particlefrontx \global\fermionfronty=\particlefronty
\ifnum\unitboxnumber > 50000
\message{   *** WARNING *** Fermion of length
\the\unitboxnumber\space requested ***   }
\ifnum\unitboxnumber > 80000
\message{   *** Reducing fermion length to 30000 (max 80000) ***   }
\global\unitboxnumber=30000 \fi \fi  
\global\fermionlength=\unitboxnumber 
\global\particleadjustx=0   \global\particleadjusty=0 
\global\numlineparts = 1    \global\numupperunits=1
\global\upperlineadjx=-200  \global\upperlineadjy=0
\global\fermionlengthx=\fermionlength    \global\fermionlengthy=\fermionlength
\gslanttest(\fermionlengthx,\fermionlengthy)  
\global\multiply\fermionlengthx by \XDIR  
\global\multiply\fermionlengthy by \YDIR  
\global\unitboxheight=\fermionlengthy   \global\unitboxwidth=\fermionlengthx   
\global\advance \fermionlengthx by \particleadjustx
\global\advance \fermionlengthy by \particleadjusty
\global\particlelengthx=\fermionlengthx
\global\particlelengthy=\fermionlengthy  
\boxlengthdefault    \rearcoords    \midcoords
\global\fermionbackx=\particlebackx     \global\fermionbacky=\particlebacky
\ifcase\LINECONFIGURATION  
\ifnum\XDIR=0 
\gdef\upperunitbox{\line(\XDIR,\YDIR){\boxlengthy}} 
\else
\gdef\upperunitbox{\line(\XDIR,\YDIR){\boxlengthx}}
\fi
\else \UNIMPERROR
\fi
}
 \fi   
\global\advance\fermioncount by 1  
\ALLfermion   
}
\gdef\selectscalar{
\ifnum\scalarcount=0 
\newcount\scalarlength
\newcount\scalarlengthx
\newcount\scalarlengthy
\newcount\scalarfrontx  
\newcount\scalarfronty  
\newcount\scalarbackx
\newcount\scalarbacky
\gdef\ALLscalar{
\global\scalarfrontx=\particlefrontx   
\global\scalarfronty=\particlefronty   
\numlineparts = 1      \numupperunits=\unitboxnumber
\ifcase\LINECONFIGURATION
\global\upperlineadjx=-200     \global\upperlineadjy=0 
\slanttest(\seglength,\gaplength)   
\gdef\upperunitbox{\line(\XDIR,\YDIR){\seglength}}
\else \UNIMPERROR 
\fi
\global\unitboxwidth=\seglength  \global\advance\unitboxwidth by \gaplength
\global\multiply \unitboxwidth by \XDIR
\global\unitboxheight=\seglength  \global\advance\unitboxheight by \gaplength
\global\multiply \unitboxheight by \YDIR
\global\particleadjustx=\gaplength \global\multiply\particleadjustx by \XDIR 
\global\particleadjusty=\gaplength \global\multiply\particleadjusty by \YDIR
\negate\particleadjustx   \negate\particleadjusty   
\setparticle  
\global\scalarlengthx=\particlelengthx  
\global\scalarlengthy=\particlelengthy  
\ifnum\boxlengthx > 50000
\message{   *** WARNING *** Scalar of length in excess of 50000cp requested!}\fi
\ifnum\boxlengthy > 50000
\message{   *** WARNING *** Scalar of length in excess of 50000cp requested!}\fi
\global\scalarbackx=\pbackx      \global\scalarbacky=\pbacky   
}
 \fi   
\global\advance\scalarcount by 1  
\ALLscalar
}
\gdef\selectphoton{   
\ifnum\photoncount=0 
\newcount\numwiggles    \newcount\numwigglespo
\global\newcount\photonlengthx
\global\newcount\photonlengthy
\global\newcount\photonfrontx  
\global\newcount\photonfronty  
\global\newcount\photonbackx
\global\newcount\photonbacky
\newcount\halfwigglelength
\global\font\Twelverom=cmr12
\global\font\Tenrom=cmr10
\gdef\Lbr{{\Twelverom(}}   \gdef\Rbr{{\Twelverom)}}
\gdef\SLbr{{\Tenrom(}}     \gdef\SRbr{{\Tenrom)}}
\gdef\Smile{{\large$\smile$}}  
\gdef\Frown{{\large$\frown$}}  
\ifdim\BIGPHOTONS>0pt  \gdef\Smile{$\smile$} \gdef\Frown{$\frown$} \fi
\gdef\selectphoton{   
\global\advance\photoncount by 1  
\global\photonfrontx=\particlefrontx   
\global\photonfronty=\particlefronty   
\ifnum\unitboxnumber > 50
\message{   *** WARNING *** Photon with 
\the\unitboxnumber\space half-wiggles requested ***   }
\ifnum\unitboxnumber > 150
\message{   *** Reducing photon length to 10 half-wiggles (max 150) ***   }
\ifnum\unitboxnumber > 1000
\message{   *** Probable Cause:  Photon selected instead of Fermion ***   }
\fi \global\unitboxnumber=10 \fi \fi  
\numwiggles=\unitboxnumber
\divide\numwiggles by 2
\global\unitboxnumberpo=\numwiggles 
\global\multiply \unitboxnumberpo by -1
\numwigglespo=\unitboxnumber
\advance\numwigglespo by \unitboxnumberpo 
\global\numlineparts = 2  
\global\numupperunits=\numwigglespo  
\global\numlowerunits=\numwiggles  
\particleadjustx=0  
\particleadjusty=0  
\ifcase\LINEDIRECTION
     \Nphoton    
\or  \NEphoton   
\or  \Ephoton    
\or  \SEphoton   
\or  \Sphoton    
\or  \SWphoton   
\or  \Wphoton    
\or  \NWphoton   
\else\DIRECTERROR \fi
\setplength
\global\divide\plengthx by 2  \global\divide\plengthy by 2
\rearcoords  \boxlengthdefault   \midcoords
\global\photonbackx=\pbackx  
\global\photonbacky=\pbacky  
\global\photonlengthx=\plengthx  
\global\photonlengthy=\plengthy  
}
\gdef\SETUNITBOX(#1)[#2][#3]{ 
\gdef\upperunitbox{\oval(#1,#1)[#2]}
\gdef\lowerunitbox{\oval(#1,#1)[#3]}
}
\gdef\Nphoton{  
\ifcase\LINECONFIGURATION  
\setcoords(-490,-250,0)(260,1250,0)[0,2000]
\gdef\upperunitbox{\SLbr}   \gdef\lowerunitbox{\SRbr}
\particleadjusty=10
\or 
\setcoords(-271,-501,0)(250,1250,0)[0,2000]   
\gdef\upperunitbox{\SRbr}   \gdef\lowerunitbox{\SLbr}
\or 
\particleadjusty=0
\setcoords(-501,-351,0)(300,1400,0)[0,2200]
\gdef\upperunitbox{\Lbr}   \gdef\lowerunitbox{\Rbr}
\or 
\setcoords(-353,-499,0)(300,1400,0)[0,2200]
\gdef\upperunitbox{\Rbr}   \gdef\lowerunitbox{\Lbr}
\or 
\setcoords(-481,-371,0)(280,1300,0)[0,2000]
\gdef\upperunitbox{\Lbr}   \gdef\lowerunitbox{\Rbr}
\particleadjusty=150
\ifnum\numwiggles=\number\numwigglespo \particleadjustx=-50 \fi
\or 
\setcoords(-321,-391,0)(280,1300,0)[0,2000]
\gdef\upperunitbox{\Rbr}   \gdef\lowerunitbox{\Lbr}
\particleadjusty=150
\ifnum\numwiggles=\number\numwigglespo \particleadjustx=80 \fi
\or 
\setcoords(-490,-260,0)(300,1500,0)[0,2400]
\gdef\upperunitbox{\Lbr}   \gdef\lowerunitbox{\Rbr}
\or 
\setcoords(-301,-531,0)(300,1500,0)[0,2400]
\gdef\upperunitbox{\Rbr}   \gdef\lowerunitbox{\Lbr}
\else \UNIMPERROR
\fi
}
\gdef\NEphoton{    
\ifcase\LINECONFIGURATION  
\setcoords(425,425,0)(1250,0,0)[1250,1250]       \SETUNITBOX(1250)[br][tl]  
\ifnum\numwigglespo > \number \numwiggles \particleadjustx=15 \fi
\or 
\setcoords(1050,-200,0)(625,625,0)[1250,1250]    \SETUNITBOX(1250)[tl][br]
\ifnum\numwigglespo > \number \numwiggles \particleadjustx=25 \fi
\or 
\setcoords(500,500,0)(1400,0,0)[1400,1400]       \SETUNITBOX(1400)[br][tl]
\or 
\setcoords(1200,-200,0)(700,700,0)[1400,1400]    \SETUNITBOX(1400)[tl][br]  
\or 
\setcoords(400,400,0)(1200,0,0)[1200,1200]       \SETUNITBOX(1200)[br][tl]  
\or 
\setcoords(1000,-200,0)(600,600,0)[1200,1200]    \SETUNITBOX(1200)[tl][br]
\else \UNIMPERROR
\fi
\numupperunits=\numwiggles   \numlowerunits=\numwigglespo
}
\gdef\Ephoton{    
\ifcase\LINECONFIGURATION  
\setcoords(-285,715,0)(-150,-400,0)[2005,0]
\gdef\upperunitbox{\Frown}   \gdef\lowerunitbox{\Smile}
\or  
\setcoords(-285,715,0)(-420,-170,0)[2005,0]
\gdef\upperunitbox{\Smile}   \gdef\lowerunitbox{\Frown}
\else \UNIMPERROR
\fi
\particleadjustx=-15 
}
\gdef\SEphoton{   
\ifcase\LINECONFIGURATION  
\setcoords(-200,1050,0)(-625,-625,0)[1250,-1250] \SETUNITBOX(1250)[tr][bl]
\ifnum\numwigglespo > \number \numwiggles \particleadjustx=25 \fi
\or 
\setcoords(425,425,0)(0,-1250,0)[1250,-1250]     \SETUNITBOX(1250)[bl][tr]
\ifnum\numwigglespo > \number \numwiggles \particleadjustx=15 \fi
\or 
\setcoords(-200,1200,0)(-700,-700,0)[1400,-1400] \SETUNITBOX(1400)[tr][bl]  
\or 
\setcoords(500,500,0)(0,-1400,0)[1400,-1400]     \SETUNITBOX(1400)[bl][tr]  
\or 
\setcoords(-200,1000,0)(-600,-600,0)[1200,-1200] \SETUNITBOX(1200)[tr][bl]
\particleadjustx=-20
\or 
\setcoords(420,420,0)(0,-1200,0)[1200,-1200]     \SETUNITBOX(1200)[bl][tr]
\particleadjustx=40
\else \UNIMPERROR
\fi
}
\gdef\Sphoton{  
\ifcase\LINECONFIGURATION  
\setcoords(-252,-490,0)(-740,-1740,0)[0,-2000]
\gdef\upperunitbox{\SRbr}   \gdef\lowerunitbox{\SLbr}
\or 
\setcoords(-490,-260,0)(-740,-1740,0)[0,-2002]
\gdef\upperunitbox{\SLbr}   \gdef\lowerunitbox{\SRbr}
\or 
\setcoords(-299,-449,0)(-870,-1970,0)[0,-2200]
\gdef\upperunitbox{\Rbr}    \gdef\lowerunitbox{\Lbr}
\particleadjusty=-95
\or 
\setcoords(-517,-371,0)(-900,-2000,0)[0,-2200]
\gdef\upperunitbox{\Lbr}    \gdef\lowerunitbox{\Rbr}
\particleadjusty=-165
\or 
\setcoords(-299,-409,0)(-885,-1905,0)[0,-2000]
\gdef\upperunitbox{\Rbr}   \gdef\lowerunitbox{\Lbr}
\particleadjustx=50     \particleadjusty=-380
\ifodd\unitboxnumber\relax\else\particleadjustx=250 \particleadjusty=-400 \fi
\or 
\setcoords(-519,-449,0)(-900,-1920,0)[0,-2000]
\gdef\upperunitbox{\Lbr}   \gdef\lowerunitbox{\Rbr}
\particleadjusty=-370
\ifodd\unitboxnumber\relax\else\particleadjustx=-240 \particleadjusty=-400 \fi
\or 
\gdef\upperunitbox{\Rbr}   \gdef\lowerunitbox{\Lbr}
\setcoords(-325,-555,0)(-900,-2100,0)[0,-2400]
\particleadjusty=-40
\or 
\setcoords(-505,-275,0)(-900,-2100,0)[0,-2400]
\gdef\upperunitbox{\Lbr}   \gdef\lowerunitbox{\Rbr}
\particleadjusty=-30  
\else \UNIMPERROR
\fi
}
\gdef\SWphoton{  
\ifcase\LINECONFIGURATION  
\setcoords(-825,-825,0)(0,-1250,0)[-1250,-1250]     \SETUNITBOX(1250)[br][tl]  
\or 
\setcoords(-175,-1425,0)(-625,-625,0)[-1250,-1250]  \SETUNITBOX(1250)[tl][br]  
\or 
\setcoords(-900,-900,0)(0,-1410,0)[-1400,-1400]     \SETUNITBOX(1400)[br][tl]  
\or 
\setcoords(-200,-1600,0)(-700,-700,0)[-1400,-1400]  \SETUNITBOX(1400)[tl][br]  
\or 
\setcoords(-800,-800,0)(0,-1200,0)[-1200,-1200]     \SETUNITBOX(1200)[br][tl]  
\or 
\setcoords(-200,-1400,0)(-600,-600,0)[-1200,-1200]  \SETUNITBOX(1200)[tl][br]  
\else \UNIMPERROR
\fi
}
\gdef\Wphoton{
\ifcase\LINECONFIGURATION 
\setcoords(-2245,-1245,0)(-150,-400,0)[-2005,0]
\gdef\upperunitbox{\Frown}   \gdef\lowerunitbox{\Smile}
\or 
\setcoords(-2245,-1245,0)(-400,-150,0)[-2005,0]
\gdef\upperunitbox{\Smile}   \gdef\lowerunitbox{\Frown}
\else \UNIMPERROR
\fi
\particleadjustx=57 
\ifnum\numwigglespo=\number\numwiggles \particleadjustx=0  \fi
\numlowerunits=\numwigglespo   \numupperunits=\numwiggles
}
\gdef\NWphoton{  
\ifcase\LINECONFIGURATION  
\setcoords(-200,-1425,0)(625,625,0)[-1250,1250]   \SETUNITBOX(1250)[bl][tr]
\or 
\setcoords(-825,-825,0)(0,1250,0)[-1250,1250]     \SETUNITBOX(1250)[tr][bl]
\ifnum\numwigglespo > \number \numwiggles \particleadjusty=-15 \fi
\or 
\setcoords(-200,-1600,0)(700,700,0)[-1400,1400]   \SETUNITBOX(1400)[bl][tr]
\or 
\setcoords(-900,-900,0)(0,1400,0)[-1400,1400]     \SETUNITBOX(1400)[tr][bl]
\or 
\setcoords(-200,-1400,0)(600,600,0)[-1200,1200]   \SETUNITBOX(1200)[bl][tr]  
\or 
\setcoords(-800,-800,0)(0,1200,0)[-1200,1200]     \SETUNITBOX(1200)[tr][bl]  
\else \UNIMPERROR
\fi
}
  \fi   
\selectphoton
}
\gdef\selectgluon{   
\ifnum\gluoncount=0 \input gluonsetup  \fi
\selectgluon
}
\gdef\selectespecial{\UNIMPERROR}
\gdef\checkvertex{ 
\ifnum\vertexcount=-1   \input VERTEX  \fi}
\gdef\drawvertex#1[#2#3](#4,#5)[#6]{\checkvertex\drawvertex#1[#2#3](#4,#5)[#6]}
\gdef\vertexcap#1{\checkvertex\vertexcap#1}
\gdef\vertexcaps{\checkvertex\vertexcaps}
\gdef\vertexlink#1{\checkvertex\vertexlink#1}
\gdef\vertexlinks{\checkvertex\vertexlinks}
\gdef\stemvertex#1{\checkvertex\stemvertex#1}
\gdef\stemvertices{\checkvertex\stemvertices}
\gdef\flipvertex{\checkvertex\flipvertex}
\global\arrowlength=349  
\gdef\drawarrow[#1#2](#3,#4){
\global\LDIR=#1
\SETDIR
\global\boxlengthx=#3  
\global\boxlengthy=#4  
\ifdim#2=1pt  
\adjx=\arrowlength      \adjy=\arrowlength
\multiply\adjx by \XDIR \multiply\adjy by \YDIR  
\slanttest(\adjx,\adjy)
\global\advance\boxlengthx by \adjx    \global\advance\boxlengthy by \adjy
\fi
\ifnum\phantomswitch=0\put(\boxlengthx,\boxlengthy){\vector(\XDIR,\YDIR){0}}\fi
}  
\gdef\SETFRONTSTEM{
\EITHERSTEM=\FRONTSTEM   \advance\EITHERSTEM by \BACKSTEM
\ifdim\EITHERSTEM>0pt
\global\stemlengthx=\stemlength   \global\stemlengthy=\stemlength   
\global\absstemlength=\stemlength   
\SETDIR
\gslanttest(\stemlengthx,\stemlengthy)
\gslanttest(\absstemlength,\REG)  
\ifnum\XDIR=0 \stemlengthx=0 \fi
\ifnum\YDIR=0 \stemlengthy=0 \fi
\global\multiply\stemlengthx by \XDIR
\global\multiply\stemlengthy by \YDIR
\ifdim\FRONTSTEM=1pt 
\ifnum\phantomswitch=0
          \put(\pfrontx,\pfronty){\line(\XDIR,\YDIR){\absstemlength}}\fi
\global\advance\plengthx by \stemlengthx
\global\advance\plengthy by \stemlengthy
\global\advance\PFRONTx by \stemlengthx   
\global\advance\PFRONTy by \stemlengthy
\global\advance\pmidx by \stemlengthx
\global\advance\pmidy by \stemlengthy
\global\advance\pbackx by \stemlengthx
\global\advance\pbacky by \stemlengthy
\ifnum\LTYPE=3
\global\photonfrontx=\PFRONTx  \global\photonfronty=\PFRONTy
\global\photonbackx=\pbackx    \global\photonbacky=\pbacky
\fi  
\ifnum\LTYPE=4
\global\gluonfrontx=\PFRONTx  \global\gluonfronty=\PFRONTy
\global\gluonbackx=\pbackx    \global\gluonbacky=\pbacky
\fi  
\fi  
\fi  
}    
\gdef\SETBACKSTEM{
\ifdim\BACKSTEM=1pt 
\ifnum\phantomswitch=0
       \put(\pbackx,\pbacky){\line(\XDIR,\YDIR){\absstemlength}}\fi
\global\advance\plengthx by \stemlengthx
\global\advance\plengthy by \stemlengthy
\global\advance\pbackx by \stemlengthx
\global\advance\pbacky by \stemlengthy
\fi  
\global\stemlength=275  \FRONTSTEM=0pt  \BACKSTEM=0pt 
}    
\gdef\drawloop#1[#2#3](#4,#5){  
\input LOOPS  
\drawloop#1[#2#3](#4,#5)}
\Feynmanlength  

\long\def\inst#1{\par\nobreak\kern 4pt\nobreak
    {\it #1}\par\vskip 10pt plus 3pt minus 3pt}

\begin{document}
{\pagestyle{empty}

\begin{flushright}
\babar-CONF-\BABARPubYear/\BABARPubNumber \\
SLAC-PUB-\SLACPubNumber
\end{flushright}

\par\vskip 3cm

\begin{center}
\Large \bf A measurement of the branching fraction of the exclusive decay
{\boldmath \bkog\ }
\end{center}
\bigskip

\begin{center}
\large The \babar\ Collaboration\\
\mbox{ }\\
\today
\end{center}
\bigskip \bigskip

\begin{center}
\large \bf Abstract
\end{center}
The $\btosgam$ transition proceeds by a loop ``penguin'' diagram. It
may be used to  measure precisely the couplings of the top quark 
and to search for the effects of any new particles appearing in the loop.
We present a preliminary measurement of the branching fraction of the exclusive
decay, $\bkog$. We use $8.6 \times 10^6$ $\BB$ decays to measure 
\brresult\ . 

\vfill
\centerline{Submitted to the XXX$^{th}$ International Conference on
High Energy Physics, Osaka, Japan.}
\newpage
}

\begin{center}
\small

The \babar\ Collaboration
\bigskip

B.~Aubert,
A.~Boucham,
D.~Boutigny,
I.~De Bonis,
J.~Favier,
J.-M.~Gaillard,
F.~Galeazzi,
A.~Jeremie,
Y.~Karyotakis,
J.~P.~Lees,
P.~Robbe,
V.~Tisserand,
K.~Zachariadou
\inst{Lab de Phys.\ des Particules, F-74941 Annecy-le-Vieux, CEDEX, France}
% ...

\end{center}\newpage

\setcounter{footnote}{0}

\begin{picture}(14000,14000)
\drawline\fermion[\E\REG](10000,1000)[4000]
\put(\pmidx,1100){b}
\drawline\scalar[\NE \REG](\fermionbackx,\fermionbacky)[3]
\put(14500,\pmidy){W}
\drawline\photon[\N \REG](\scalarbackx,\scalarbacky)[6]
\put(17000,\pmidy){$\gamma$}
\drawline\scalar[\SE \REG](\scalarbackx,\scalarbacky)[3]
\put(21000,\pmidy){W}
\drawline\fermion[\E\REG](\scalarbackx,\scalarbacky)[4000]
\put(\pmidx,1100){s}
\drawline\fermion[\W\REG](\scalarbackx,\scalarbacky)[8000]
\put(17000,1100){u,c,t}
\end{picture}
\begin{figure}[!htb]
\begin{center}
\vspace{0.1in}
\caption{The leading order Feynman diagram for the $\bkog$
decay}
\label{fig:feynman}
\end{center}
\end{figure}

In the Standard Model the exclusive decay $\Bz \to \Kstarz \gamma$ proceeds 
by the $\btosgam$ loop ``penguin'' diagram shown in Fig.~\ref{fig:feynman}. 
Precise measurements of decay modes involving these transitions and modes
with the related $\btodgam$  transition such as $\brog$ will allow
measurements of the top quark couplings $V_{ts}$ and $V_{td}$.
The strength of these transitions may also be  enhanced by the
presence of non-Standard Model contributions~\cite{joanne}. In minimal
supersymmetric models, for example, the $W$ can be replaced by a charged
Higgs leading to significant enhancements in the branching fraction.
  In the first year of running the \babar\ experiment has  accumulated
a dataset comparable to the world's largest to date, and  
this will increase by an order of magnitude over the next few years.
The large dataset available and the state of the art detection systems
employed at $\babar$ will allow measurements and searches of unprecedented
precision. A comprehensive program to study these decays is now underway.
The first step in this program is the preliminary measurement of the 
branching fraction of the exclusive decay mode $\bkog$ using the 
leading decay mode, \kpi. 
Here \Kstarz\ refers to the $K^{*0}(892)$ resonance, and charge 
conjugate channels are assumed throughout. The most precise measurement of 
the branching fraction to date, 
$\BR(\bkog)=(4.55^{+0.72}_{-0.68} \pm 0.34) \times 10^{-5}$, 
is from the CLEO collaboration~\cite{CLEO} and is in agreement with Standard
Model predictions of $(3.3-6.3) \times 10^{-5}$~\cite{theory}.

The data were collected with the \babar\ detector
\cite{babar} at the \pep2\ asymmetric \epem\ storage ring~\cite{pep}. 
The results presented in this paper are based upon an
integrated luminosity of 7.5\invfb\ of  data corresponding to
$8.6 \times 10^6$
\BB\ meson pairs recorded at the $\Y4S$ energy (``on-resonance'') and
1.1\invfb\ below the $\Y4S$ energy (``off-resonance''). The
\babar\ detector simulation is based upon GEANT~\cite{geant} and tuned with
data.  Events taken from random triggers are used to measure the beam 
backgrounds. They  are mixed into the simulated events, so that small changes 
in detector conditions, including dead channels, are also simulated.
The simulated events are processed in the same manner as data.

The selection criteria for this analysis have been optimized to maximize
$S^{2}/(S+B)$ where $S$ is the number of signal candidates expected, assuming
$\BR(\bkog) = 4.55 \times 10^{-5}$, and $B$ is the expected number
of background candidates determined from Monte Carlo. We compute quantities
in both the laboratory frame and the rest frame of the \Y4S. Quantities
computed in the rest frame are denoted by an asterisk; e.g. $\ebeam$ is the
energy of the \ep\ and \en\ beams which are symmetric in the \Y4S\ rest frame.

We begin the selection by requiring a high energy photon candidate in
the calorimeter. The calorimeter
consists of 6580 thallium doped CsI crystals  
arranged in a non-projective barrel and 
forward endcap geometry. We require the photon candidate to have an
energy, measured in the laboratory frame, between 1.5 and 4.5\gev\ 
and further require  $2.20 < \egcms < 2.85$\gev. A photon candidate is 
defined as a calorimeter cluster~\cite{babar} in a
region of good calorimetry ($-0.74 <\cos{\theta} < 0.93$), where $\theta$
is the polar angle to the beam axis. The cluster must be 
isolated from any track. The distribution of individual crystal 
energies in the cluster 
is required to have only one maximum and a lateral profile consistent with
a photon shower. These requirements remove backgrounds from high energy
$\piz$ and $\eta$ mesons where the two photons from the decay have merged into 
one cluster. In addition we form the combination of the high energy 
photon candidate with all other photons in the event with energy
greater than 50 (250)\mev\ and require 
the invariant mass of the combination not to lie within two standard deviations
of the known $\piz(\eta)$ mass.
    
We next reconstruct the $\Kstarz$ from $\Kp$ and $\pim$
candidates, by considering all pairs of
tracks in the event. The tracks  are required to be well reconstructed in
the tracking detectors and to originate from a vertex consistent with the
$\epem$ interaction point. The tracking detector consists of a five layer
double-sided silicon vertex tracker plus a forty layer stereo-axial 
drift chamber.  A track is identified as a kaon if it
is projected to pass through the fiducial volume of the particle identification
detector. This detector is a novel ring imaging Cherenkov detector (DIRC) in 
which the radiating medium, an elongated quartz bar, also acts as a light guide
to project the light cone onto an array of photo-multiplier tubes at the
backward end of the detector. The high light yield and low mass of this 
detector are a significant advance in particle identification
over previous experiments. We require a cone of Cherenkov light consistent 
in time and angle with a kaon of the measured track momentum. 
The $\Kstarz$ reconstruction is completed by requiring
the invariant mass of the candidate pairs to be within 90\mevcc\ of
the $\Kstarz$ mass: $ 806 < \mkpi < 986$\mevcc.

\begin{figure}[!htb]
\begin{center}
\includegraphics[height=7cm]{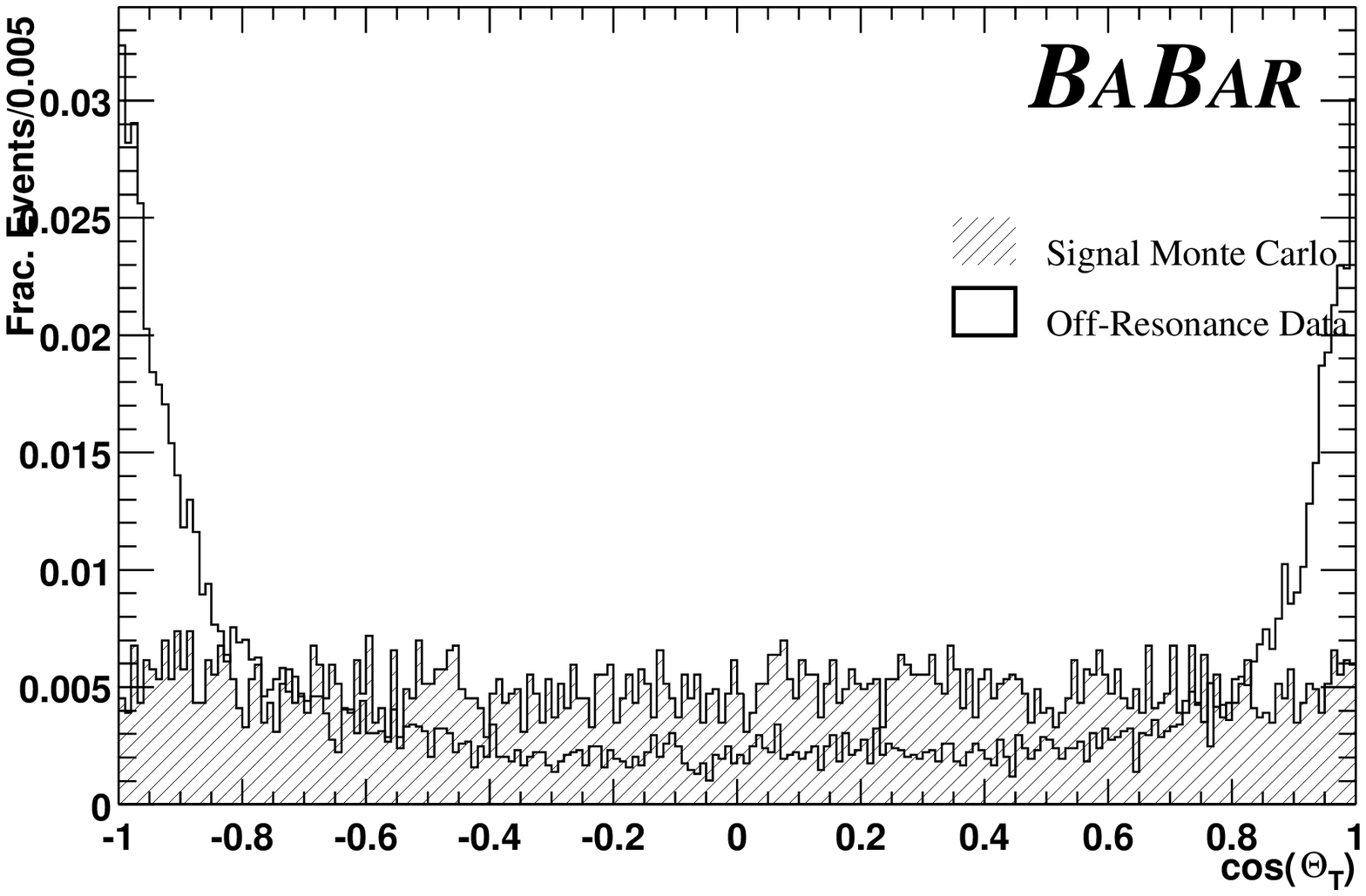}
\caption{The event shape variable \ctg\ for \bkog,  \kpi\ Monte Carlo and
off resonance data.}
\label{fig:costhr}
\end{center}
\end{figure}

The \Bz\ candidates are reconstructed from the \Kstarz\ and $\gamma$
candidates. There are  backgrounds from continuum \qqbar\ production with the
high energy photon originating  from  initial state radiation or from a
$\piz$ or $\eta$. We exploit event topology differences between signal
and background to reduce the continuum contribution. The thrust 
vector $\vec{T}^{*}$ of the event is computed, 
excluding the \Bz\ daughter candidates. 
Fig.~\ref{fig:costhr} shows the distribution of \ctg\ for
signal Monte Carlo events and off resonance data, where \tg\ is the angle 
between the high energy photon candidate and $\vec{T}^{*}$.
In this frame the \BB\ pairs are
produced approximately at rest and therefore decay isotropically with a
flat distribution in $\ctg$. The
\qqbar\ pair is produced above threshold recoiling against each other in
a jet-like topology, which results in a \ctg\ distribution peaking at
$\pm 1$ . 
We require $|\ctg| < 0.78$. Backgrounds are further suppressed
by constructing two additional event shape variables which are uncorrelated 
to \ctg. The angle of the \Bz\ candidate direction with respect to the 
beam axis, $\theta^{*}_{B}$ follows a $1-\cos^{2}\theta_{B}$ distribution 
and is flat for  \qqbar\ production. We require $|\ctb| < 0.76$. 
The helicity of the $\Kstarz$  decay, $\theta^{*}_{H}$, is defined as 
the angle of the \Kp\ in the \Kstarz\ rest frame  with respect to the 
flight direction of the \Kstarz\ in the \FourS\ rest frame.  
This follows a $1-\cos^{2}\theta^{*}_{H}$ distribution in \bkog, \kpi, 
whereas the \qqbar\  background is flat. We require $|\cth| < 0.7$. 

Since the \Bz\ mesons are produced via $\epem \to \Y4S \to \BB$, the energy of
the \Bz\ is given by the beam energy, \ebeam. The beam energy
is measured much more precisely than the energy of the \Bz\ candidate 
daughter particles. In particular for the decay $\bkog$ the energy 
resolution of
the high energy photon dominates the measured \Bz\ candidate energy. We 
reconstruct the \Bz\ candidate substituting  \ebeam\ for the
measured energy of the candidate daughters.  We define the difference
of the  beam energy and energy of the \Bz\ daughters, 
$\de = E^{*}_{K^*} + \egcms - \ebeam$. The \Bz\ mass is given by
$\mes= \sqrt{{\ebeam}^2-|p_{B}^{*2}|}$, where $|p^{*}_{B}|$ is the momentum
of the \Bz\ candidate calculated using the measured momenta of the charged
daughters and the energy of the photon. In the calculation of \mes\ we
rescale the measured photon energy, \egcms\ by a factor $\kappa$ so
that $\ebeam = E^{*}_{K^*}+\kappa\egcms$. This procedure corrects
for the low energy tail in the $\egcms$ distribution due to the incomplete 
containment of showers in the calorimeter.
Using \mes\ is an order of magnitude more precise
than using the invariant mass. In addition the rescaling of \egcms\ 
enhances the resolution on \mes\ by 20\%. The resolution of 3.0\mevcc\ 
is dominated by the beam energy spread. 
We select candidates with $\mes >5.2$\gevcc.

\begin{figure}[!htb]
\begin{center}
\includegraphics[height=7cm]{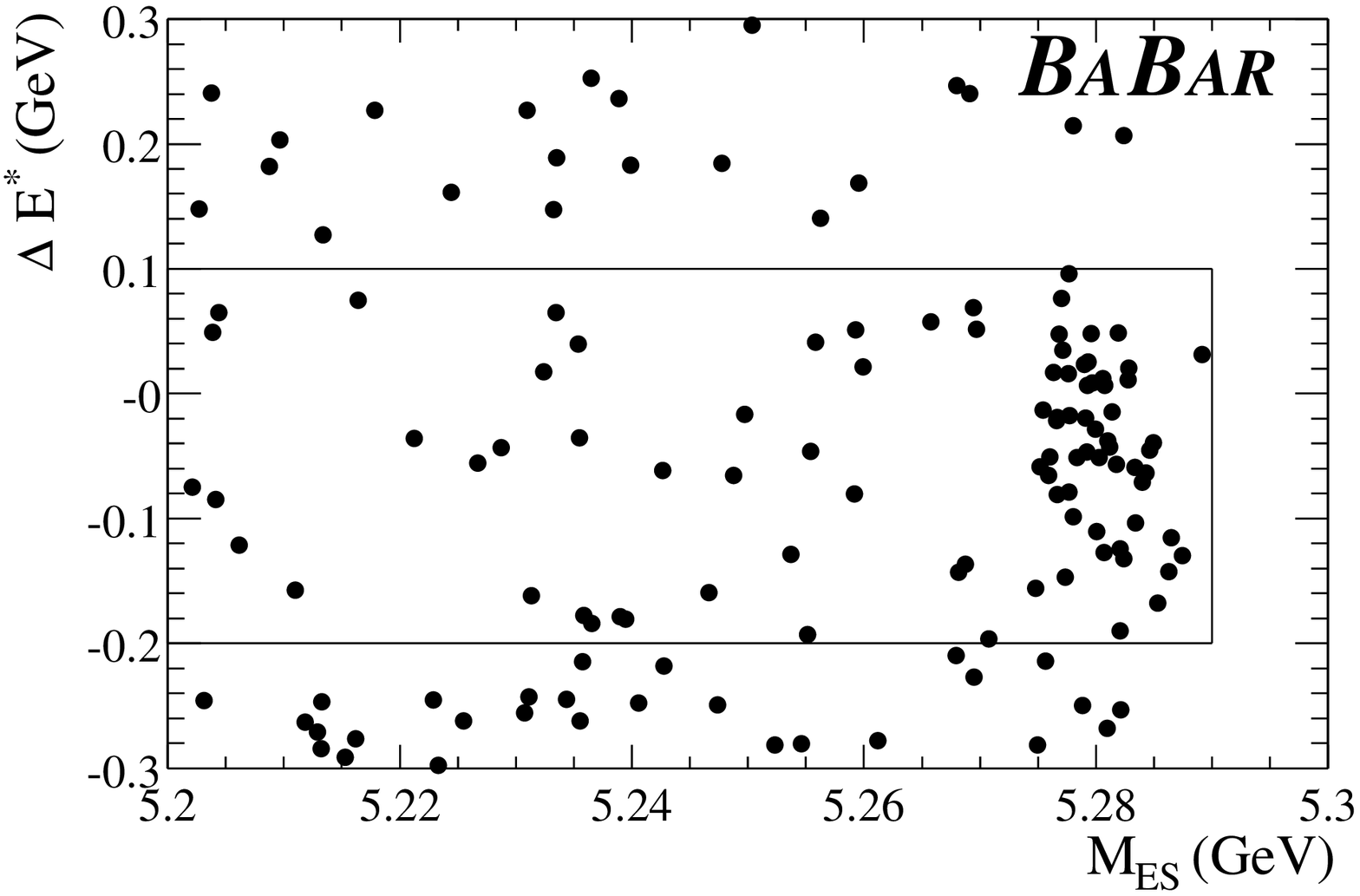}
\caption{\de\ versus \mes\  for \bkog,\kpi\ candidates from \lumi\ \BB\ decays.
The box indicates the boundaries used to make the projection of the 
\de--\mes\ plane onto the \mes\ axis in Fig.~\ref{fig:datamb}.} 
\label{fig:datademb}
\end{center}
\end{figure}

\begin{figure}[!htb]
\begin{center}
\includegraphics[height=7cm]{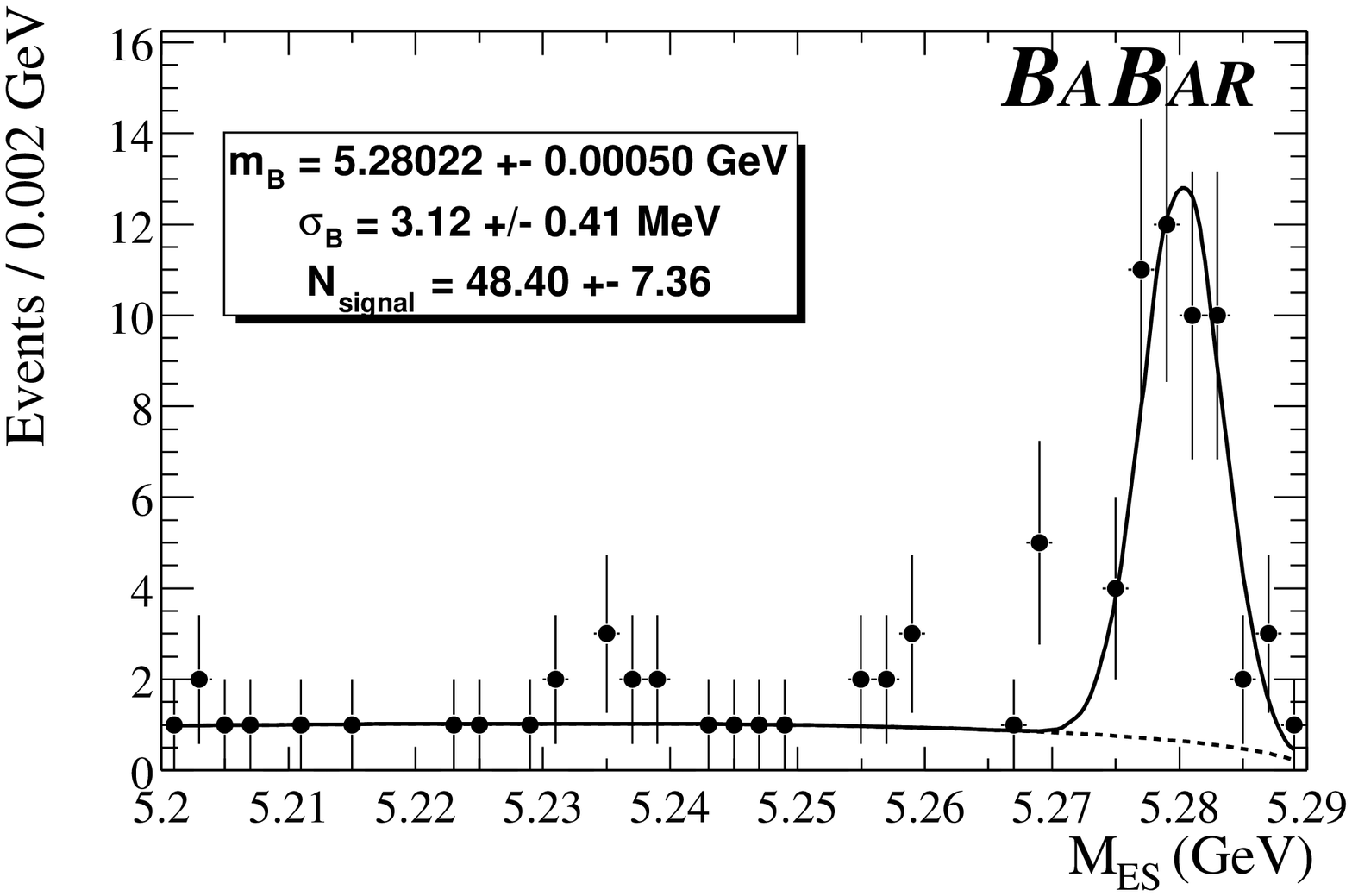}
\caption{The \mes\ projection  for \bkog, \kpi\ candidates from \lumi\ 
\BB\ decays. The fit uses a parameterized background function, described 
in the text, and a Gaussian for the signal.}
\label{fig:datamb}
\end{center}
\end{figure}

Figure~\ref{fig:datademb} shows the $\de$ versus \mes\ distribution for 
on-resonance data. Figure~\ref{fig:datamb} shows
the projection onto the \mes\ axis, requiring   $ -200 < \de < 100$\mev. 
The \mes\ distribution is fitted 
using an unbinned maximum likelihood fit. The background is empirically 
described by a function~\cite{argusf}:
\[
\frac{dN}{d\mes} \propto \mes . \sqrt{1-\frac{\mes^{2}}{{\ebeam}^2}}
            . \exp\left[-\zeta\left(1-\frac{\mes^{2}}{{\ebeam}^2}\right)\right]
\]
where the parameter $\zeta$ is measured from the off-resonance data sample
which is required to pass the same selection
criteria as the on-resonance data sample except that we remove the $\cth$ cut, 
and relax the selection requirements to $|\ctg| < 0.95$ and $|\de| < 500$\mev\ 
to gain statistics. We measure $\zeta = 22 \pm 13$. The fit to the
on-resonance data uses this fixed value of $\zeta$ and adds a signal Gaussian
whose mean and width are allowed to vary. 
We find a signal of $48.4 \pm 7.3$ events with the error coming from 
the statistical error of the fit.  

\begin{figure}[!htb]
\begin{center}
\includegraphics[height=7cm]{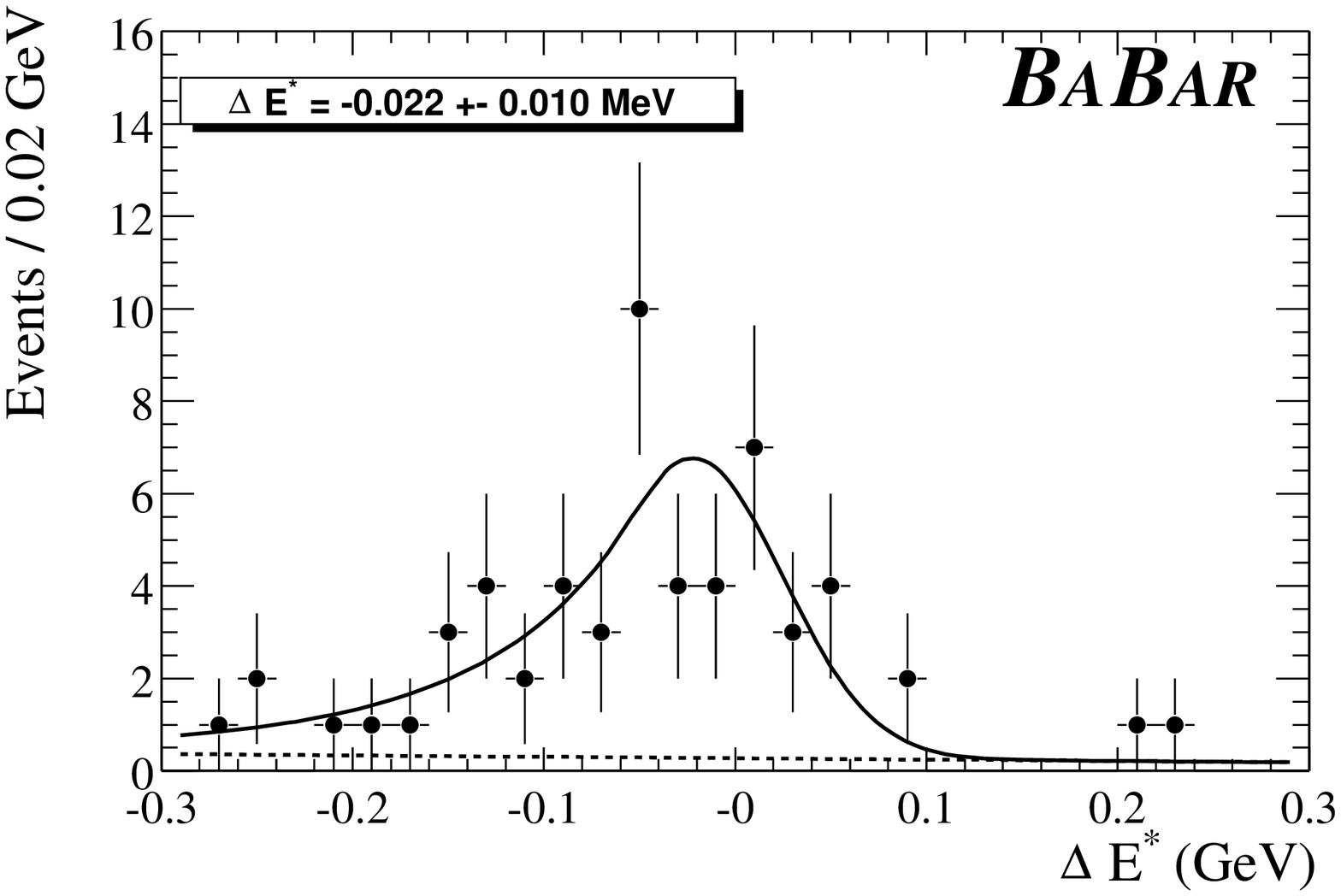}
\caption{The \de\ projection for \bkog, \kpi\ candidates from 
\lumi\ \BB\ decays.}
\label{fig:datade}
\end{center}
\end{figure}

\begin{figure}[!htb]
\begin{center}
\includegraphics[height=7cm]{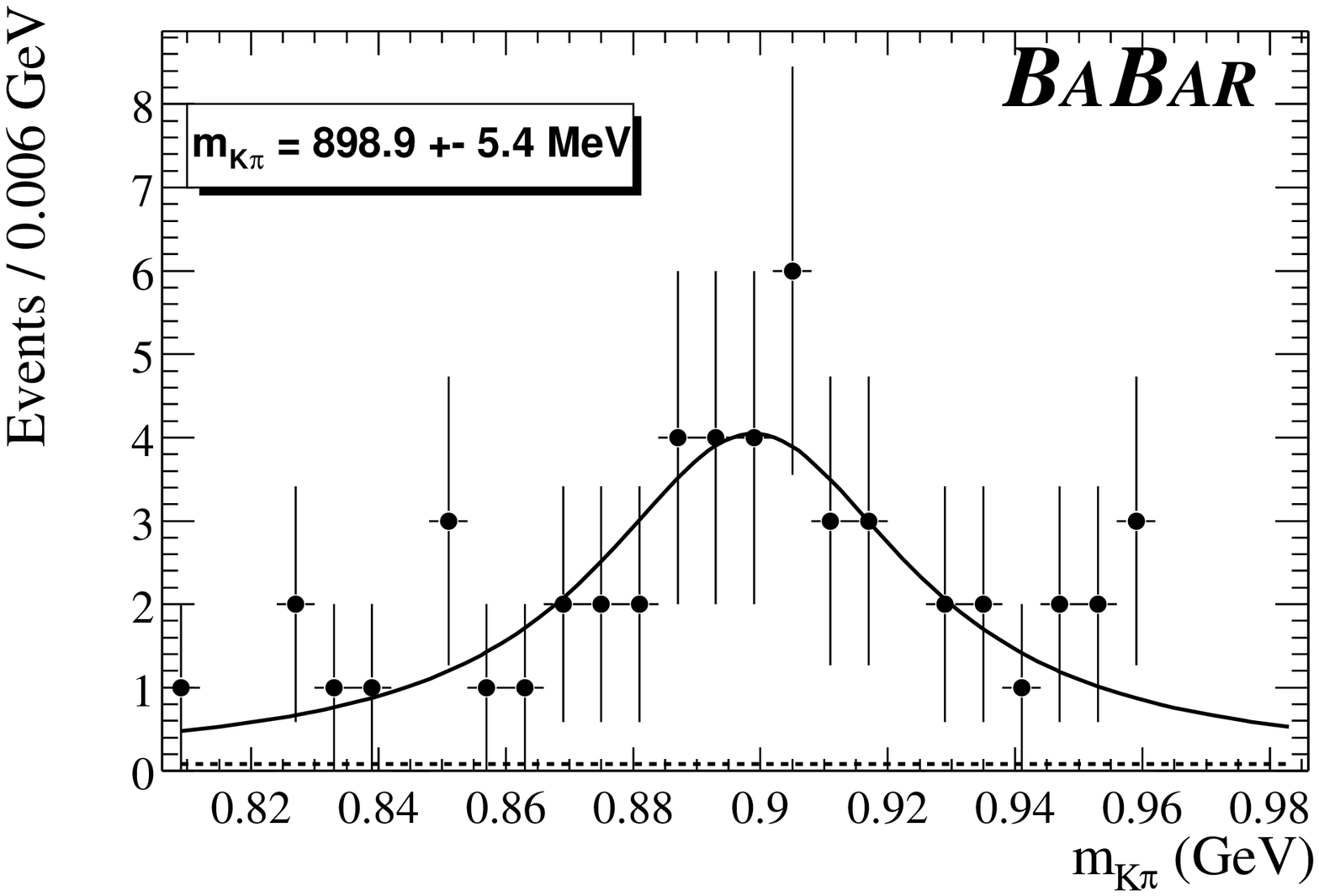}
\caption{The \mkpi\ distribution for \bkog\ , \kpi\ candidates from \lumi\ 
\BB\ decays. The fit is to a Breit-Wigner shape}
\label{fig:datakstar}
\end{center}
\end{figure}

As a consistency check we plot in Fig.~\ref{fig:datade} the \de\ projection
by requiring $ 5.274 < \mes < 5.285$\gevcc.  
The \de\ distribution is fitted using a fixed shape determined
from the Monte Carlo sample which is in good agreement with the data.
We also plot \mkpi\ in Fig.~\ref{fig:datakstar} by 
requiring $ 5.274 < \mes < 5.285$\gevcc\ and 
$ -200 < \de < 100$\mev. We fit using a Breit-Wigner shape and determine
that the signal is consistent with coming from a $\Kstarz$. 

The efficiency for the selection of $\bkog$ candidates is $(15.5\pm 0.3)\%$.
The branching fraction is determined using the yield, the efficiency
and the total number of \BB\ events in the sample. The number of
\BB\ events is determined by counting the number of events passing
a loose generic hadron requirement and subtracting the non-\BB\ component
estimated by scaling off-resonance data. We measure the number of \BB\
events in the sample to be \lumi.
The  branching fraction is measured to be  \brresult\  consistent both with 
previous measurements and with the Standard Model expectations. The
first error is the statistical uncertainty in the yield. The second
error is the systematic uncertainty.

%
The total systematic error of $8.6\%$ is a quadratic sum  of several 
uncorrelated components tabulated in Table~\ref{tab:systematic}. The
systematic uncertainty in the background shape is obtained from varying 
the parameter $\zeta$ within the allowed range from the off-resonance data.

The primary check on track efficiency is obtained by
studying the probability for observing drift chamber versus silicon detector-only tracks in
inclusive $\Dz\to\Km\pip\pip\pim$. This is compared with the rate for finding the third track 
in one-versus-three topology
tau-pair decays. A final check is the observed multiplicity distribution in \FourS\ events.
A systematic error of 2.5\% per track
with $\pt>1$\gevc\ on the overall efficiency scale is determined by comparing the three methods.
In addition we measure
the tracking resolution with $\epem \to \mu^{+} \mu^{-}$ events and
compare to Monte Carlo expectations. We find a difference in the tail
of this distribution which leads to a systematic uncertainty in the $\Kstarz$ selection efficiency. 

The kaon identification efficiency in the DIRC is derived from a sample of 
$D^{*+} \to D^{0}\pip$ in which the charge of the associated slow pion tags 
the flavor of the D meson and hence the charge of the kaon in the 
subsequent $D^{0} \to \Km \pip$. 
The errors on this measured value are taken as a systematic uncertainty in the efficiency
for kaon identification.

The calorimeter energy resolution 
is measured in data using $\piz$ and $\eta$ meson decays, and  
$\epem \to \epem \gamma$ events. These are compared to Monte Carlo simulated 
events of the same processes. The differences result in a systematic 
uncertainty in the energy resolution
which can broaden the \de\ distribution causing an uncertainty in the
efficiency of the \de\ requirement. An overall energy scale uncertainty 
is estimated by using a data sample of $\eta$ meson decays with 
symmetric energy photons. The deviation in the reconstructed $\eta$ 
mass from the nominal
$\eta$ mass estimates the  uncertainty in the measured single photon
energy. This resultant deviation in the central value of the \de\
distribution causes a systematic uncertainty in the efficiency of the
\de\ requirement. The photon efficiency uncertainty is estimated using a 
sample of \eerad\ events compared to Monte Carlo simulated \eerad\ 
events. The expected energy and position of the photon can be derived from a 
fit to the beam energy and the measured track momenta  for the $\epem$.
We apply the photon selection cuts and compare the efficiency in the 
data and the Monte Carlo sample to derive the systematic uncertainty.
The $\piz/\eta$ veto efficiency is tested by ``embedding'' a
Monte Carlo generated photon into both off-resonance data and
off-resonance Monte Carlo events. The difference in the measured efficiency is used as
the systematic uncertainty. The merged $\piz$ modeling is tested
by comparing a sample of merged $\piz$ in \taurho\ data and 
Monte Carlo. We assign a systematic in the photon selection efficiency
by comparing the fraction of merged $\piz$ removed in the two samples.

\begin{table}[!htb]
\caption{The fractional systematic uncertainties in the measurement of
\BR(\bkog).}
\begin{center}
\begin{tabular}{|c|c|} \hline
Uncertainty                         & \% of \BR(\bkog) \\ \hline\hline
Tracking efficiency                 &  5.0            \\
$B$ counting                        &  3.6            \\
Kaon identification efficiency      &  3.0            \\
Track resolution                    &  3.0            \\
Calorimeter energy resolution       &  2.5            \\
Background shape                    &  2.3            \\
Monte Carlo Statistics              &  1.9            \\
Calorimeter energy scale            &  1.0            \\
Calorimeter efficiency              &  1.0            \\
\piz/$\eta$ veto                   &  1.0            \\
Merged \piz\ modeling               &  1.0            \\ \hline\hline
Total                               &  8.6            \\ \hline
\end{tabular}
\end{center}
\label{tab:systematic}
\end{table}

In conclusion, we report a preliminary measurement of the branching
fraction of the rare decay \brresult\ with a precision comparable to  
previous experiments and consistent with Standard Model expectations. 


\section*{Acknowledgments}
\label{sec:Acknowledgments}

We are grateful for the contributions of our \pep2\ colleagues in
achieving the excellent luminosity and machine conditions
that have made this work possible.
We acknowledge support from the
Natural Sciences and Engineering Research Council (Canada),
Institute of High Energy Physics (China),
Commissariat \`a l'Energie Atomique and
Institut National de Physique Nucl\'eaire et de Physique des Particules
(France),
Bundesministerium f\"ur Bildung und Forschung
(Germany),
Istituto Nazionale di Fisica Nucleare (Italy),
The Research Council of Norway,
Ministry of Science and Technology of the Russian Federation,
Particle Physics and Astronomy Research Council (United Kingdom), the
Department of Energy (US),
and the National Science Foundation (US). In addition, individual support 
has been received from the Swiss 
National Foundation, the A. P. Sloan Foundation, the Research Corporation,
and the Alexander von Humboldt Foundation.
The visiting groups wish to thank 
SLAC for the support and kind hospitality
extended to them.

\end{document}